\documentclass[doublecol]{epl2}
\usepackage{bm}
\usepackage{graphicx}
\usepackage{amsmath,amssymb}
\usepackage{dcolumn}

\title{Nonvanishing anisotropic magnetoresistance in Rashba two-dimensional electron systems with nonmagnetic disorders}
\shorttitle{Nonvanishing anisotropic magnetoresistance} %Insert here a short version of the title if it exceeds 70 characters

\author{C. M. Wang\thanks{E-mail:\email{cmwangsjtu@gmail.com}}\and M. Q. Pang}
\shortauthor{C. M. Wang \etal}

\institute{
  School of Physics and
Electrical Engineering, Anyang Normal University, Anyang 455000,
China}




\abstract{We study anisotropic magnetoresistance (AMR) in a
spin-polarized two-dimensional electron gas with Rashba spin-orbit
coupling and nonmagnetic disorder collision. We show that AMR
exists, arising from the combined effect of in-plane magnetization,
spin-orbit coupling, and nonmagnetic remote disorder scattering.
Further, numerical evaluation demonstrates that the smoothness of
the remote disorder can strongly affect AMR, and this AMR is
sensitive to the electron density. Large magnitude of AMR
($\approx24\%$) is obtained for low density system with strong
spin-orbit splitting.}

\begin{document}

\maketitle

\section{Introduction}

Magnetotransport phenomena in ferromagnetic semiconductor, such as
anisotropic magnetoresistance
(AMR)\cite{Thomson546,smit1951,ieee,jaoul1977,Baxter2002,rushforth1938amc,kato2008iam,rushforth2009oac,Tang2003,Pappert2007,Lim2006,kovalev},
have attracted significant attentions due to applications in the
emerging field of spintronics \cite{wolf2001rmc,Jungwirth2006}.
AMRs, including longitudinal and transverse AMRs, are the response
of magnetoresistance to the relative angle between magnetization and
current in magnetic materials. Both the longitudinal and transverse
conductivities show the symmetric feature: $\sigma_{xx}(\bm
M_0)=\sigma_{xx}(-\bm M_0)$ and $\sigma_{yx}(\bm
M_0)=\sigma_{yx}(-\bm M_0)$, where the magnetization $\bm M_0$ is
usually in the two-dimensional plane ($x$-$y$ plane). However, one
should note that for ordinary charge Hall effect (including
anomalous Hall effect \cite{ahe}) the transverse conductivity obeys
the antisymmetric relation, $\sigma_{yx}(\bm M_0)=-\sigma_{yx}(-\bm
M_0)$. Here $\bm M_0$ is normal to the plane.

Experimentally, AMR has been extensively studied in diluted magnetic
semiconductors recently. Rushforth {\it et al.} investigated the
physical origin of the noncrystalline and crystalline components of
AMR in diluted magnetic semiconductors \cite{rushforth1938amc}. Shin
{\it et al.} explored the temperature dependence of AMR in
ferromagnetic (Ga,Mn)As films \cite{Shin2007}. A giant transverse
AMR was also observed in this ternary ferromagnetic semiconductor
(Ga,Mn)As \cite{Pappert2007,Tang2003}. In contrast to the extensive
experimental studies of AMR, the theoretical interpretation of AMR
is relatively poor. The experimental analyses are usually based on a
phenomenological treatment \cite{smit1951,ieee}. The full Boltzmann
theory simulations have been made to study the origin of the sources
of AMR in $p$-type magnetic semiconductor
\cite{rushforth1938amc,rushforth2009oac,Karel}. Kato. {\it et al.}
analyzed the intrinsic AMR in spin-polarized two-dimensional gas
(2DEG) with Rashba spin-orbit coupling (SOC) \cite{kato2008iam}.
They showed that AMR vanishes unless the relaxation time is
spin-related. Recently, Trushin {\it et al.} studied AMR for Rashba
or Dresselhaus spin-orbit splitting electron system with polarized
magnetic impurities \cite{Trushin2009}. In the above theoretical
studies, the microscopic mechanism of AMR is considered as due to
the anisotropic carriers lifetime, arising from the combined effect
of the SOC and the scattering by the polarized magnetic impurities.
The spin-dependent scattering is the essential factor in AMR
\cite{kovalev,rushforth1938amc,rushforth2009oac,kato2008iam,Trushin2009,vborn}.
In most studies, the nonmagnetic disorder potential is taken as
$\delta$-form type. However in realistic heterostructure, the
electron density is not large
  enough to screen the nonmagnetic impurities, where the interaction between
  electron and disorder is long-ranged. Hence, the effect of
  electron-impurity scattering on AMR is far from being understood, completely.

In this paper, we employ the kinetic equation approach to
investigate AMR in two-dimensional electron system in the presence
of Rashba-type spin-orbit interaction and in-plane magnetization. We
show that the combined effect of SOC, in-plane magnetization, and
nonmagnetic long-range impurity can lead to AMR. At the same time,
numerical evaluation demonstrates the disorder-distance- and
electron-density-related feature of AMR. The present study may
provide another mechanism of AMR in spin-orbit interaction 2DEG with
in-plane magnetization. The nonmagnetic wave-vector-dependent
disorders, coupling to the SOC and magnetization, induces the
magnetotransport anisotropy.

\section{Theoretical approach}

We consider a 2DEG confined in a [001]-grown III-V semiconductor
heterostructure with Rashba SOC and a homogenous in-plane
magnetization $\bm M_0$. The $x$ and $y$ axes are taken along [100]
and [010] direction, respectively. Hence, the noninteracting
one-particle Hamiltonian can be written as
\begin{equation}\label{ham}
 \check{H}=\frac{k^2}{2m}+\alpha(\hat z\times \bm \sigma)\cdot{\bm
k}-M_x\sigma_x-M_y\sigma_y.
\end{equation}
Here $m$ is the electron effective mass, $\bm
\sigma\equiv(\sigma_x,\sigma_y,\sigma_z)$ are the Pauli matrices,
${\bm k}\equiv(k\cos\theta_{\bm k},k\sin\theta_{\bm k})$ is the
two-dimensional electron wave vector, $\alpha$ is the Rashba SOC
parameter, and $\bm M\equiv(M_x,M_y)=M(\cos\xi,\sin\xi)=g\mu_B\bm
M_0\equiv g\mu_BM_0(\cos\xi,\sin\xi)$ with ${g}$ as the effective
$g$-factor, $\mu_B$ as
    the Bohr magneton, and $\xi$ as the angle between the magnetization $\bm M_0$ and [100]-axis.

The above Hamiltonian \eqref{ham} can be diagonalized into
$\hat{H}=U_{\bm k}^\dag \check{H}U_{\bm k}={\rm
diag}[\varepsilon_{1}(\bm k),\varepsilon_{2}(\bm k)]$ in the
helicity basis with the help of the following local unitary
transformation
\begin{equation}\label{}
U_{\bm k}=\frac{1}{\sqrt{2}}\left(
                              \begin{array}{cc}
                                1 & 1 \\
                                ie^{i\chi_{\bm k}} & -ie^{i\chi_{\bm k}} \\
                              \end{array}
                            \right).
\end{equation}
Here the energy dispersion $\varepsilon_{\mu}(\bm
k)=\frac{k^2}{2m}+(-1)^\mu\varepsilon_{\rm RM}(\bm k)$ with
$\varepsilon_{\rm RM}(\bm k)=\sqrt{\alpha^2k^2+M^2+2\alpha
kM\sin(\xi-\theta_{\bm k})}$, $\mu=1,2$ as the helix band index, and
\begin{equation}\label{}
\chi_{\bm k}=\tan^{-1}\frac{\alpha k\sin\theta_{\bm
k}-M\cos\xi}{\alpha k\cos\theta_{\bm k}+M\sin\xi}.
\end{equation}

Now we consider the quasi-two-dimensional system is driven by a weak
dc electric field $\bm E$ along [100] direction. Obviously, in order
to carry out the evaluation of the AMR, it is necessary to determine
the matrix electron distribution function. The kinetic equation for
the $2\times2$ matrix distribution function $\rho(\bm k)$ in the
stationary linear
 response regime can
be
 derived, where the elastic electron-impurity scattering is taken
into account in the self-consistent Born approximation
\cite{liu195329,liu2005des,lin2006she}. Following the procedure of
these papers, the distribution
 function can be obtained as, $\rho(\bm k)=\rho^{(0)}(\bm k)+\rho^{(1)}(\bm k)+\rho^{(2)}(\bm
 k)$, with equilibrium distribution function $\rho^{(0)}(\bm k)={\rm diag}\big\{n_{\rm F}[\varepsilon_{1}(\bm k)],n_{\rm F}[\varepsilon_{2}(\bm k)]\big\}$, and $n_{\rm F}(x)$ as the Fermi-Dirac function. Here
$\rho^{(1)}(\bm k)$ and $\rho^{(2)}(\bm k)$ are
 collision-unrelated and collision-related matrix distribution functions in
 the first order of electric field, respectively. The
 collision-unrelated distribution function $\rho^{(1)}(\bm k)$ is off-diagonal matrix with the
 elements given by
 \begin{equation}\label{}
 \rho^{(1)}_{12}(\bm k)=\rho^{(1)}_{21}(\bm k)=\frac{eE_0}{4\varepsilon_{\rm RM}}\frac{\partial \chi_{\bm k}}
 {\partial k_x}\big\{n_{\rm F}[\varepsilon_{1}(\bm k)]-n_{\rm F}[\varepsilon_{2}(\bm k)]\big\},
 \end{equation}
where $E_0$ is the strength of the electric field. This distribution
function is associated with the interband transition between two
spin-orbit-coupled bands, making no contribution to charge
conductivity. However, it is important for spin Hall effect, which
results in the collision-independent intrinsic spin Hall effect
\cite{ShuichiMurakami09052003,sinova2004uis,lin2006she}. The
collision-related distribution function $\rho^{(2)}(\bm k)$ is
determined by the coupled equations
\begin{align}
  eE_0\frac{\partial n_{\rm F}(\varepsilon_{\bm k\mu})}{\partial k_x}&=\pi\sum_{\bm q\mu'}\vert V(\bm k-\bm q)\vert^2
  \Omega_{\mu\mu'}\nonumber\\\times&\left[\rho_{\mu\mu}^{(2)}(\bm k)-
  \rho_{\mu'\mu'}^{(2)}(\bm q)\right]\delta\big[\varepsilon_{\mu}(\bm k)-\varepsilon_{\mu'}(\bm q)\big], \label{eq1}\\
  4\varepsilon_{\rm RM}(\bm k){\rm Re}\rho_{12}^{(2)}(\bm k)&=\pi\sum_{\bm q\mu\mu'}\vert V(\bm k-\bm q)\vert^2
  \bar{\Omega}_{\mu\mu'}\nonumber\\\times&\left[\rho_{\mu\mu}^{(2)}(\bm k)-\rho_{\mu'\mu'}^{(2)}(\bm q)
  \right]\delta\big[\varepsilon_{\mu}(\bm k)-\varepsilon_{\mu'}(\bm q)\big].\label{eq2}
\end{align}
Here $\Omega_{\mu\mu'}=1+(-1)^{\mu+\mu'}\cos(\chi_{\bm k}-\chi_{\bm
q})$ and $\bar{\Omega}_{\mu\mu'}=(-1)^{\mu'}\sin(\chi_{\bm
k}-\chi_{\bm q})$. $V(\bm k-\bm q)$ is the nonmagnetic impurity
scattering potential. ${\rm Re}\rho_{12}^{(2)}(\bm k)$ represents
the real part of the off-diagonal distribution function
$\rho_{12}^{(2)}(\bm k)$. One should note that here the weak
scattering limit is assumed, where we restrict ourselves to the
leading order of the impurity concentration. In this case, the
imaginary part of the off-diagonal distribution function
$\rho_{12}^{(2)}(\bm k)$ can be ignored completely \cite{liu195329}.
In the above kinetic equations, both the interband and the intraband
transitions are considered.

In order to study AMR, it is necessary to evaluate the drift
velocity. In spin basis, the two in-plane matrix velocity operators
read
\begin{equation}\label{}
\check{v}_x=\left(
           \begin{array}{cc}
             \frac{k_x}{m} & i\alpha \\
             -i\alpha & \frac{k_x}{m} \\
           \end{array}
         \right),
\end{equation}
\begin{equation}\label{}
 \check{v}_y=\left(
           \begin{array}{cc}
             \frac{k_y}{m} & \alpha \\
             \alpha & \frac{k_y}{m} \\
           \end{array}
         \right).
\end{equation}
It is clear that the velocity operators in spin basis are
independent of the magnetization. Moreover, the expressions of the
velocity operators are the same as the ones of semiconductor
heterostructure with Rashba spin-orbit interaction in the absence of
magnetization. However, in the helicity basis, the single-particle
operators of velocity $\hat v_i=U_{\bm k}^\dag \check v_iU_{\bm k}$
($i=x,y$), rely on the magnetization through the energy spectrum and
the angle $\chi_{\bm k}$, and are given by
\begin{equation}\label{}
\hat{v}_x=\left(
           \begin{array}{cc}
             \frac{\partial \varepsilon_{1}(\bm k)}{\partial k_x} & i\alpha\sin\chi_{\bm k} \\
             -i\alpha\sin\chi_{\bm k} & \frac{\partial \varepsilon_{2}(\bm k)}{\partial k_x}\\
           \end{array}
         \right),
\end{equation}
\begin{equation}\label{}
 \hat{v}_y=\left(
           \begin{array}{cc}
             \frac{\partial \varepsilon_{1}(\bm k)}{\partial k_y} & -i\alpha\cos\chi_{\bm k} \\
             i\alpha\cos\chi_{\bm k} & \frac{\partial \varepsilon_{2}(\bm k)}{\partial k_y} \\
           \end{array}
         \right).
\end{equation}
One find that the off-diagonal elements of in-plane velocity
operators are also nonvanishing in helicity basis. The corresponding
macroscopical drift velocities are obtained by taking the
statistical average over them, $v_i=\frac{1}{N}\sum_{\bm k}{\rm
Tr}[\rho(\bm k)\hat v_i]$, and expressed as
\begin{equation}
  v_i = \frac{1}{N}\sum_{\bm k\mu}\frac{\partial \varepsilon_\mu(\bm k)}{\partial k_i}\rho_{\mu\mu}^{(2)}(\bm
  k).\label{vel}
\end{equation}
Here $N$ is the electron density. It can be seen that the average
velocities only depend on the diagonal element of velocity
operators. One should emphasize that in clean limit approximation,
the imaginary part of the off-diagonal element of collision-related
distribution function vanishes. Hence, the drift velocities only
relate to the diagonal elements of velocity operators and the
diagonal elements of distribution function. And the expressions of
average velocities become the same as the usual form of two band
system without interband coupling. We only need Eq. \eqref{eq1} to
determine the diagonal elements of distribution function. The real
part of off-diagonal elements, ${\rm Re}\rho_{12}^{(2)}(\bm k)$, is
essential for calculation of spin Hall effect
\cite{liu2005des,lin2006she} and anomalous Hall effect
\cite{liu195329}. The longitudinal and transverse conductivities are
defined by $\sigma_{xx}=Nev_x/E_0$ and $\sigma_{yx}=Nev_y/E_0$,
respectively.

For AMR, one can find that the longitudinal and transverse
conductivities obey the symmetric relations: $\sigma_{xx}(\bm
M_0)=\sigma_{xx}(-\bm M_0)$ and $\sigma_{yx}(\bm
M_0)=\sigma_{yx}(-\bm M_0)$. We can understand these properties as
follows: The eigenenergy $\varepsilon_\mu(\bm M_0,\bm k)$ and angle
$\chi_{\bm k}(\bm M_0)$ satisfy $\varepsilon_\mu(-\bm M_0,-\bm
k)=\varepsilon_\mu(\bm M_0,\bm k)$ and $\chi_{-\bm k}(-\bm
M_0)=\pi+\chi_{\bm k}(\bm M_0)$. Therefore distribution function
satisfies $\rho_{\mu\mu}^{(2)}(-\bm M_0,-\bm
  k)=-\rho_{\mu\mu}^{(2)}(\bm M_0,\bm
  k)$. Note that for brevity, the argument $\bm M_0$ for eigenenergy,
  distribution function and so on, is dropped elsewhere.
  When $\bm M_0\rightarrow-\bm M_0$, we make transformation
  $\bm k\rightarrow-\bm k$ in Eq. (\ref{vel}). This transformation will
not change the total integral, hence the conductivities satisfy the
symmetric property. This property is in vivid contrast to the one of
anomalous Hall effect, where the transverse conductivity obeys the
antisymmetric relation. Now we remark these relations from the point
of view of the time reversal symmetry. For AMR, the conductivities
are related to the diagonal elements of distribution function, which
are proportional to a momentum-dependent effective transport
relaxation time $\tau_{\rm tr}$ (see footnote\footnote{Actually, the
diagonal elements of distribution function are proportional to the
quantity with dimension of time, relating to several band-dependent
relaxation times \cite{liu195329}. We call this quantity as
effective transport relaxation time.}). However, the anomalous Hall
conductivity relies on the off-diagonal elements. The off-diagonal
elements do not depend on this effective transport relaxation time,
directly, and the disorder plays only an intermediate role
\cite{liu195329}. Under time reversal, for AMR, $\sigma_{xx}(-\bm
M_0,-\tau_{\rm tr})=-\sigma_{xx}(\bm M_0,\tau_{\rm tr})$ and
$\sigma_{yx}(-\bm M_0,-\tau_{\rm tr})=-\sigma_{yx}(\bm M_0,\tau_{\rm
tr})$. By considering the relations between the conductivities and
the effective relaxation time, one can obtain the symmetric feature.
While for anomalous Hall conductivity $\sigma_{yx}(-\bm
M_0)=-\sigma_{yx}(\bm M_0)$ under time reversal, and the
antisymmetric relation is obtained directly.

\section{Numerical results}
We numerically investigate the combined effect of Rashba SOC,
magnetization, and long-range nonmagnetic electron-impurity
scattering on the AMR in InAs/InSb heterostructure. The long-range
electron-impurity collision is considered as due to remote
nonmagnetic charged impurities separated at a distance $s$ from the
interface. The potential takes the form $V({ q})=U({q})/\kappa({
q})$ with $|U({ q})|^2=n_i\left( \frac{e^2}{2\epsilon_0\kappa q}
\right)^2 e^{-2qs}I(q)^2$ \cite{Stern,lei1985tdb}. Here $I(q)$ is
the form factor and $\kappa({ q})$ is the
  factor related to the Coulomb screening, the expressions
  of which can be found in Ref. \cite{lei1985tdb}. $n_i$ is
  the density of nonmagnetic remote impurity. In the calculation the electron effective mass is taken as
$m=0.04m_{\rm e}$ with $m_{\rm e}$ as the free electron
 mass. The dielectric
 constant of InAs $\kappa=15.15$. In the numerical analysis, we have
 assumed that $\varepsilon_{\rm F}>M$, {\it i.e.} both the minority
 and the majority
 bands are occupied, with $\varepsilon_{\rm F}$ as the Fermi energy.

\begin{figure}[t]
    \begin{center}
        \includegraphics[width=0.3\textwidth]{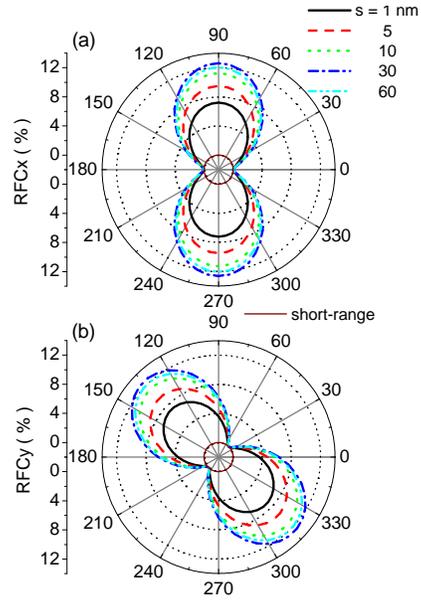}
    \end{center}
    \caption{The relative fractional change in the longitudinal (a) and
    transverse (b) conductivities as functions of the angle $\xi$ between magnetization
in the plane and [100]-axis for various remote impurity distances.
The thin wine lines are obtained for $\delta$-form short-range
electron-disorder collision. The electron density
$N=1.0\times10^{11} \,{\rm cm}^{-2}$. Rashba constant
$\alpha=3.0\times 10^{-11}\,{\rm eVm}$ and the magnetization
$M=2\,{\rm meV}$.}
    \label{fig:rfcs}
\end{figure}

\begin{figure}[t]
    \begin{center}
        \includegraphics[width=0.3\textwidth]{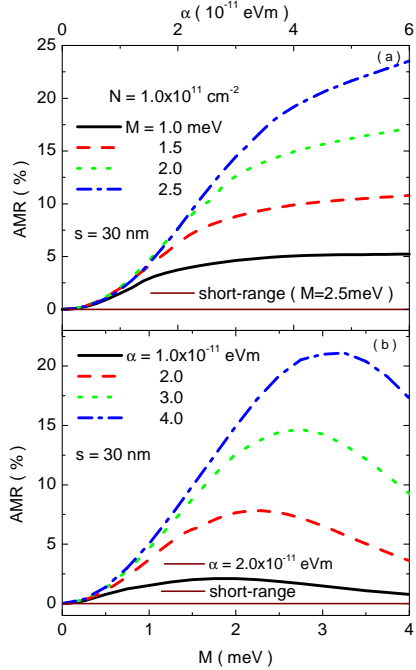}
    \end{center}
    \caption{AMR is shown as functions of spin-orbit interaction parameter
    for various magnetizations (a) and as functions of magnetization for
    various Rashba coupling parameters (b) at fixed distance $s= 30\,{\rm nm}$.
     Here the electron concentration $N=1.0\times10^{11} {\rm cm}^{-2}$. The thin wine lines indicate the corresponding AMR for short-range
electron-impurity scattering.}
    \label{fig:amr}
\end{figure}

\begin{figure}[t]
    \begin{center}
        \includegraphics[width=0.35\textwidth]{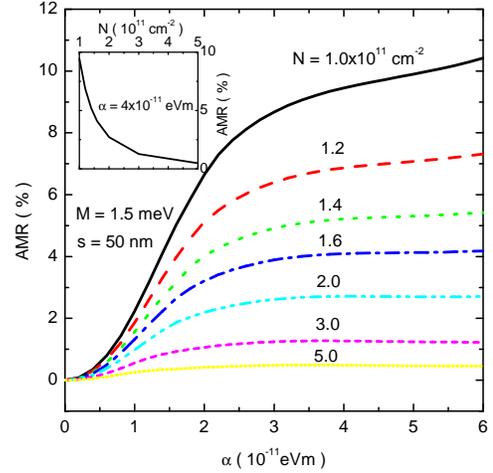}
    \end{center}
    \caption{Dependencies of AMR
     on SOC constant $\alpha$ for different electron densities. The magnetization $M=1.5\,{\rm
meV}$ and the remote impurity distance $s= 50\,{\rm nm}$. The inset
shows AMR as functions of electron density when Rashba spin-orbit
splitting $\alpha=4.0\times 10^{-11}\,{\rm eVm}$.}
    \label{fig:amrN}
\end{figure}

We first consider the relative fractional changes in the
conductivity (RFC), which are defined by
\begin{eqnarray}
  {\rm RFCx} &=& \frac{\Delta\sigma_{xx}}{\sigma_{av}}-\left(\frac{\Delta\sigma_{xx}}{\sigma_{av}}\right)_{\rm min}, \\
  {\rm RFCy} &=& \frac{\Delta\sigma_{yx}}{\sigma_{av}}-\left(\frac{\Delta\sigma_{yx}}{\sigma_{av}}\right)_{\rm
  min},
\end{eqnarray}
where $\Delta\sigma_{xx}=\sigma_{xx}-\sigma_{av}$,
$\Delta\sigma_{yx}=\sigma_{yx}-\sigma_{av}$, and $\sigma_{av}$ is
the average value of the longitudinal conductivity as the
magnetization is rotated through $360^\circ$ with respect to
[100]-axis. The subscript ``min" means the corresponding minimum
value of the fractional change. It is seen that RFC is independent
of impurity density.

In Fig. \ref{fig:rfcs}, the relative fractional changes in the
longitudinal and
    transverse conductivities are shown as functions of the angle between magnetization
in the plane and [100]-axis for this nonmagnetic remote disorder.
The corresponding thin wine solid line is obtained for $\delta$
shape short-range electron-disorder collision, $V(q)=V_0$,
independent of momentum. It is clear that the longitudinal
conductivity shows the strong anisotropy for magnetization aligned
along various direction when the nonmagnetic disorder is
long-ranged. However, the anisotropy completely vanishes for
short-range electron-impurity scattering, in agreement with the
previous studies \cite{kato2008iam,Trushin2009}. The degree of this
anisotropy depends strongly on the smoothness of the remote
disorder. With the rise of the impurity distance, the degree of the
anisotropy first enhances, and then drops when the distance is large
enough. In Fig. \ref{fig:rfcs} (b), it is seen that the transverse
conductivity also indicates the anisotropy for various direction of
magnetization. The remote disorders can affect the degree of the
anisotropy of transverse conductivity, similar to the longitudinal
conductivity. It is also found that, when the nonmagnetic disorder
becomes short-ranged, the anisotropy of transverse conductivity also
vanishes completely. This confirms that the combined effect of
Rashba SOC, in-plane magnetization, and nonmagnetic remote disorder
could lead to AMR. Our numerical evaluation shows that these
longitudinal and transverse AMRs are consistent with the standard
phenomenology due to symmetry arguments
\cite{rushforth1938amc,Trushin2009}:
${\Delta\sigma_{xx}}/{\sigma_{av}} = C_I \cos 2\xi $,
${\Delta\sigma_{yx}}/{\sigma_{av}} = C_I \sin 2\xi$. Here $C_I$ is a
dimensionless constant and is sometimes called noncrystalline
coefficient in literatures. One note that crystalline AMR
coefficient vanishes in this case, which is a special property of
the Rashba model. It is not valid for systems with other SOC, such
as Dresselhaus SOC \cite{Trushin2009}.

Now we limit ourselves to AMRs defined as the relative change
between longitudinal resistivities for magnetization along and
normal to the current direction. We take the current direction along
[100]-axis. In this case, it is found that the transverse
conductivity vanishes. Hence, AMR is given by
\begin{equation}\label{}
{\rm
AMR}=\frac{\rho_{xx}^{\parallel}-\rho_{xx}^{\perp}}{(\rho_{xx}^{\parallel}+\rho_{xx}^{\perp})/2}
=2\frac{\sigma_{xx}^{\perp}-\sigma_{xx}^{\parallel}}{\sigma_{xx}^{\parallel}+\sigma_{xx}^{\perp}}.
\end{equation}
Here $\sigma_{xx}^{\parallel}$ and $\sigma_{xx}^{\perp}$ are the
corresponding longitudinal conductivities for $\bm M\parallel\bm J$
and $\bm M\perp\bm J$ with $\bm J$ as the current density. Also one
find that this definition of AMR is disorder-density-independent.

In Fig. \ref{fig:amr}(a), we plot AMRs as functions of spin-orbit
interaction constant
    for various magnetizations at fixed impurity distance $s= 30\,{\rm
    nm}$. It is seen that the magnitude of magnetization can affect
    AMR strongly. A large AMR ($\sim24\%$) can be observed for Rashba coupling parameter up
    to $6\times10^{11}\,{\rm eVm}$ at $M=2.5\,{\rm meV}$. With the
    increment of Rashba SOC coefficient, AMR ascends and may
    saturate at strong coupling. We also evaluate the dependencies
    of AMR on the magnitude of magnetization for various SOC constants in Fig. \ref{fig:amr}(b). With the rise of
    magnetization, AMR first increases, and then decreases.
    However, AMR is always positive. The strength of
    the spin-orbit interaction can affect both the value and the
    position of maximum AMR. For short-range
    nonmagnetic impurity, AMR vanishes completely.

In order to investigate the density-related feature of AMR, in Fig.
\ref{fig:amrN}, AMRs are calculated for various electron density for
fixed magnetization and disorder distance. It is clear that AMR is
very sensitive to the electron concentration. With the increasing
density, AMR, arising from the electric remote scattering in SOC
semiconductor with in-plane magnetization, drops quickly. For
$N=5.0\times10^{11}\,{\rm cm}^{-2}$ at large coupling constant,
${\rm AMR}\sim0.45\%$. It is small but still measurable
experimentally \cite{rushforth1938amc}. The inset shows the
dependence of AMR on electron density for $\alpha=4\times
10^{-11}\,{\rm eVm}$. This implies vanishing AMR in the limit of
$k_{+}(\theta_{\bm k})\approx k_{-}(\theta_{\bm k})$ with
$k_{\pm}(\theta_{\bm k})$ as the two angle-dependent Fermi wave
vectors.

We now make some comments on the experiments to confirm the present
results. Since we deals with AMR arising from nonmagnetic disorder,
the nonmagnetic $n$-type InAs-based heterojunction can be well
satisfied. The in-plane magnetization may be induced by an in-plane
magnetic field. The magnitude of magnetic field corresponds to a
magnetization $M=1\,{\rm meV}$ is $2.16\,\rm T$ (in InAs-based
heterojunction, the effective $g$-factor $g=8$ \cite{Smith1987}).
The Rashba SOC constant can be tuned by controlling the gate
voltage. The usual Hall setup is well satisfied to measure this AMR.
We note that Papadakis {\it et al.} reported the observation of AMR
of two-dimensional holes in nonmagnetic GaAs \cite{Papadakis2000}.
Our present study may provide a possible interpretation of that
novel phenomenon. However, careful theoretical investigation should
be made for $p$-type semiconductor systems.

\section{Conclusion}
In summary, AMR for two-dimensional electron systems with a
Rashba-type spin-orbit splitting and an in-plane magnetization is
investigated for nonmagnetic impurity collision. It is found that
the combined effect of SOC, in-plane magnetization, and electric
remote disorder leads to AMR. The disorder distance can strongly
affect the degree of anisotropy of longitudinal and transverse
conductivity. The strong density-related character of AMR is also
demonstrated.

\begin{acknowledgments}
CMW thanks M. Trushin for useful discussions. CMW acknowledges
support from Research Startup Funds of AYNU. %SYL gratefully
%acknowledges support from the National Science Foundation of China
%(grant 10804073), and from the Program for New Century Excellent
%Talents in University.
\end{acknowledgments}

\end{document}